\documentclass[%
superscriptaddress,
amsmath,amssymb,
aps, prf, longbibliography,
notitlepage
]{revtex4-1}
\usepackage[hidelinks=true]{hyperref} 
\hypersetup{
  colorlinks   = true, 
  urlcolor     = blue, 
  linkcolor    = blue, 
  citecolor   = red 
}

\usepackage{graphicx}
\usepackage{dcolumn}
\usepackage{bm}
\usepackage{color}
\usepackage{pstricks}

\begin{document}


\title{Solid-on-solid contact in a sphere-wall collision in a viscous fluid}

\author{Sumit Kumar Birwa}
\email{birwasumit@gmail.com}
\affiliation{TIFR Centre for Interdisciplinary Sciences, 21 Brundavan Colony, Narsingi, Hyderabad 500075, India}
\affiliation{International Centre for Theoretical Sciences, TIFR, Shivakote, Bengaluru 560089, India}
\author{G. Rajalakshmi}
\affiliation{TIFR Centre for Interdisciplinary Sciences, 21 Brundavan Colony, Narsingi, Hyderabad 500075, India}
\author{Rama Govindarajan}
\affiliation{TIFR Centre for Interdisciplinary Sciences, 21 Brundavan Colony, Narsingi, Hyderabad 500075, India}
\affiliation{International Centre for Theoretical Sciences, TIFR, Shivakote, Bengaluru 560089, India}
\author{Narayanan Menon}
\affiliation{TIFR Centre for Interdisciplinary Sciences, 21 Brundavan Colony, Narsingi, Hyderabad 500075, India}
\affiliation{Department of Physics, University of Massachusetts, Amherst MA 01002, USA}

\date{\today}

\begin{abstract}
We study experimentally the collision between a sphere falling through a viscous fluid, and a solid plate below. It is known that there is a well-defined threshold Stokes number above which the sphere rebounds from such a collision. Our experiment tests for direct contact between the colliding bodies, and contrary to prior theoretical predictions, shows that solid-on-solid contact occurs even for Stokes numbers just above the threshold for rebounding. The dissipation is fluid-dominated, though details of the contact mechanics depend on the surface and bulk properties of the solids. Our experiments and a model calculation indicate that mechanical contact between the two colliding objects is generic and will occur for any realistic surface roughness. 


\end{abstract}

\keywords{Suggested keywords}
                             
\pacs{Valid PACS appear here}
\maketitle
\section{\label{sec:intro} Introduction}

One of the celebrated triumphs of fluid mechanics is Reynolds's explanation of how a sheared viscous fluid generates sufficient pressure to separate the moving rotors in a journal bearing \cite{reynolds1886theory}. And yet, a spoon clangs against the side of a cup as we stir a fluid. The circumstances under which collisions between solid particles occur in viscous fluids are relevant to many phenomena such as sedimentation, filtration, suspension flows, smoke and fog formation by aerosols.

In this article, we consider the simple situation of a sphere falling towards a plane in a viscous fluid. Balls rebound when bounced on the floor. However, if the air is replaced by a highly viscous fluid a ball can settle without bouncing. We ask whether solid-on-solid impact occurs during bouncing collisions, and more generally whether solid dissipation plays a role in determining the transition from bouncing to settling. 

The collision between two smooth spheres was studied by Davis \textit{et al.} \cite{davis1986elastohydrodynamic} within an elastohydrodynamic calculation. Working within the lubrication approximation, they suggested that pressure in the thin fluid film between the spheres is large enough to elastically deform them. The stored elastic strain energy is released as kinetic energy causing rebound of the particles without solid-on-solid contact. This idea must be reconciled with our everyday experience with acoustic emission and dents when solids collide within fluids.

It is now recognised \cite{davis1986elastohydrodynamic,gondret1999experiments,gondret2002bouncing,joseph2001particle,zenit1999mechanics} that the Stokes number $St$ controls the dynamics of such a collision. It compares a particle's inertia to viscous forces and is defined as
\begin{equation}
 St \equiv \frac{1}{9}\frac{\rho_s UD}{\mu} = \frac{1}{9}\frac{\rho_s}{\rho_f}Re,
\end{equation}
where $U$, $D$, $\rho_s$ and $Re$ respectively are the velocity, diameter, density and the Reynolds number of the sphere, and $\rho_f$ and $\mu$ respectively are the density and dynamic viscosity of the fluid.  
Elastohydrodynamic lubrication theory \cite{davis1986elastohydrodynamic} predicts a critical Stokes number $St_c$, below which a collision between smooth spheres does not result in a bounce. 

Experiments on sphere-wall collisions \cite{gondret1999experiments,gondret2002bouncing,joseph2001particle,zenit1999mechanics,barnocky1988elastohydrodynamic} find that bouncing dynamics do indeed collapse when plotted as a function of $St$, and that there is a transition from bouncing to settling at $St_c$ ranging from about 8 to 15. 
These experiments measure by video-imaging the coefficient of restitution, which is the ratio of the velocity just after impact to the velocity just before impact. There is only a modest variation of $St_c$ with material \cite{gondret2002bouncing,zenit1999mechanics} e.g. Gondret \textit{et al.} \cite{gondret2002bouncing} found similar $St_c$ for teflon (Young's modulus, $E=0.5$ GPa) and tungsten carbide ($E=534$ GPa).

Surface roughness can play a role when the separation between the solids becomes comparable to roughness. Theoretically, Davis \textit{et al.} \cite{davis1987elastohydrodynamic,barnocky1988elastohydrodynamic} accounted for this by implementing an inelastic collision at a cut-off distance set by the roughness. Joseph \textit{et al.} \cite{joseph2001particle} used spheres with well-characterized roughness and argued that the scatter in their data could be explained by surface roughness. Short-range forces such as Van der Waals forces could conceivably play a role close to contact, but are not considered in these pictures of bouncing.

\begin{figure*}
\begin{center}
\includegraphics[scale=0.5]{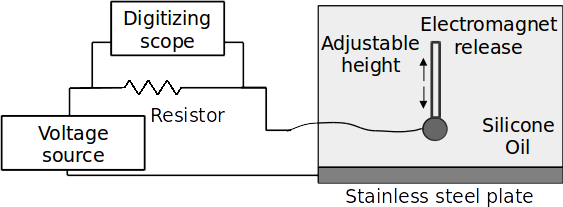}

(a)\vspace{0.5cm}

\includegraphics[scale=0.6]{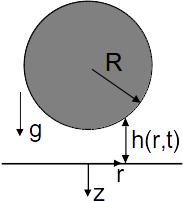}

(b)

\caption{\label{fig:setup}(a) Schematic diagram of experimental set-up. The voltage is measured across a resistor ($R_s$) of 220 k$\Omega$. (b) A sphere approaching a plane, where $r, z$ are the radial and vertical directions, $D=2R$ is the diameter of the sphere, $g$ is the acceleration due to gravity, and $h(r,t)$ is the time-dependent distance between sphere and the 
bottom wall.}
\end{center}
\end{figure*}

Whether contact occurs during rebound is still not established. It is difficult with video techniques to resolve contact dynamics spatially or temporally. In our experiments we directly address the existence and influence of solid-on-solid contact using an electrical set-up to investigate kinematics very close to the moment of impact.

\section{Experiment}

We dropped stainless steel spheres on a stainless steel plate of $10$ cm x $10$ cm x $1$ cm thickness through silicone oil of dynamic viscosity $\mu= 346 \pm 20$ kg-m/s$^2$, and density $\rho_f=970$ kg/m$^3$. As shown in Figure~\ref{fig:setup}a, a voltage was applied between the plate and the sphere.  When the sphere makes or breaks electrical contact with the plate, the circuit closes or opens \cite{king2011inelastic}. The current through the resistor was sampled by a digital oscilloscope (Tektronics DPO 4054B) at a rate of $1$ to $5$ MHz. Both direct current (DC) and alternating current (AC) voltage sources were used in the experiments.
\begin{figure}
\begin{center}
\includegraphics[scale=0.4]{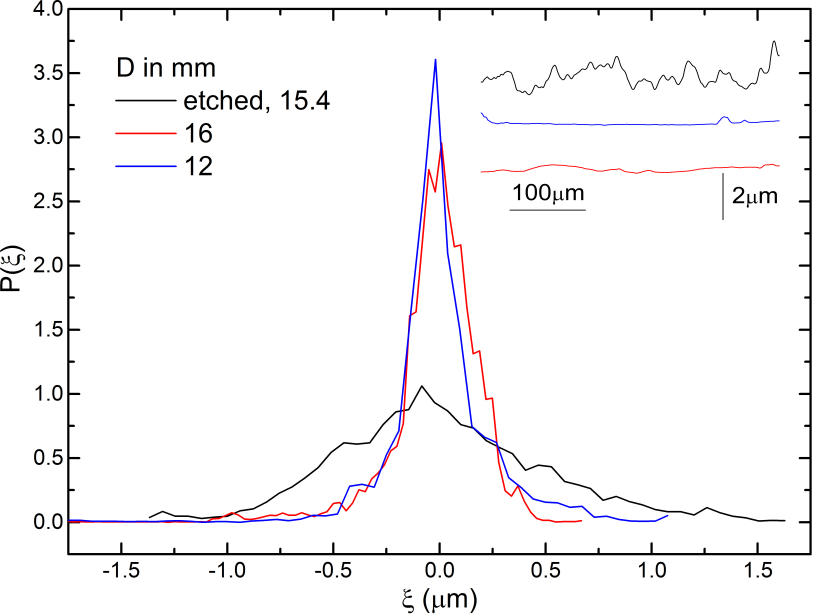}
\caption{\label{fig:roughness}Probability density $P(\xi)$ of roughness of spheres as measured with a contact profilometer. To obtain the roughness $\xi$, the path traced by profilometer as a function of distance has been subtracted from the gross curvature of the sphere. The inset shows typical profiles $\xi(s)$ of roughness as a function of the position $s$ along the surface, for all the three spheres used.}
\end{center}
\end{figure}

An electromagnet mounted on a micrometer translation stage released spheres from different initial heights, allowing us to vary the incoming velocity and the Stokes number. The release heights ranged from 1 mm to 70 mm, and were determined to an accuracy of 20 $\mu$m, corresponding to a maximum error of 2\% at the lowest height. In each case, the nominal $St$ was determined from the velocity $U$ computed at the height of the plate's surface in the absence of the plate. For this, we solved the equation of motion of the ball under gravity, buoyancy and viscous drag (Figure~\ref{fig:setup}b) using an empirical formula \cite{flagan1988fundamentals} applicable to our experimental range of Reynolds number ($Re=5.7$ to 32), 
\begin{equation}
 \rho_s V \frac{d U}{d t} = (\rho_s - \rho_f)Vg - 3\pi \mu D U (1 + 0.15 Re^{0.687}),
\end{equation}
where $V$ is the volume of the sphere. The resulting Stokes numbers ranged from 5 to 28 with a maximum error of 6.5\% at the largest $St$ (including contributions from temperature dependence of viscosity and precision of release height).

We used two types of spheres. One was as-purchased stainless steel ball bearings with density $\rho_s=7630$ kg/m$^3$ in two diameters, $D=16$ mm and 12 mm. The second type of sphere, $D=15.4$ mm, was produced by etching the 16 mm balls  with HNO$_3$ in a 1:3 aqueous solution for 10 minutes. We measured surface topography with a Dektak contact profilometer and extracted position-dependent roughness from the difference $\xi$ between the measured height, and the best-fit spherical profile. The as-purchased spheres had smooth patches with low rms roughness around $\sim 0.025$ $\mu$m interspersed with widely separated pits (of typical height $1-2$ $\mu$m) and mounds (of typical height $0.25$ $\mu$m), with lateral size $\sim 10$ $\mu$m. These numbers varied slightly between individual spheres. The etched spheres had larger, but more uniform roughness with an rms value $\sim 0.4$ $\mu$m but several larger peaks of the order of $1-2$ $\mu$m (more details are in appendix A). The typical lateral scale of the roughness is $\sim100$ $\mu$m. 

\section{Results}

\begin{figure}
\begin{center}
\includegraphics[width=6.225cm,height=5.25cm]{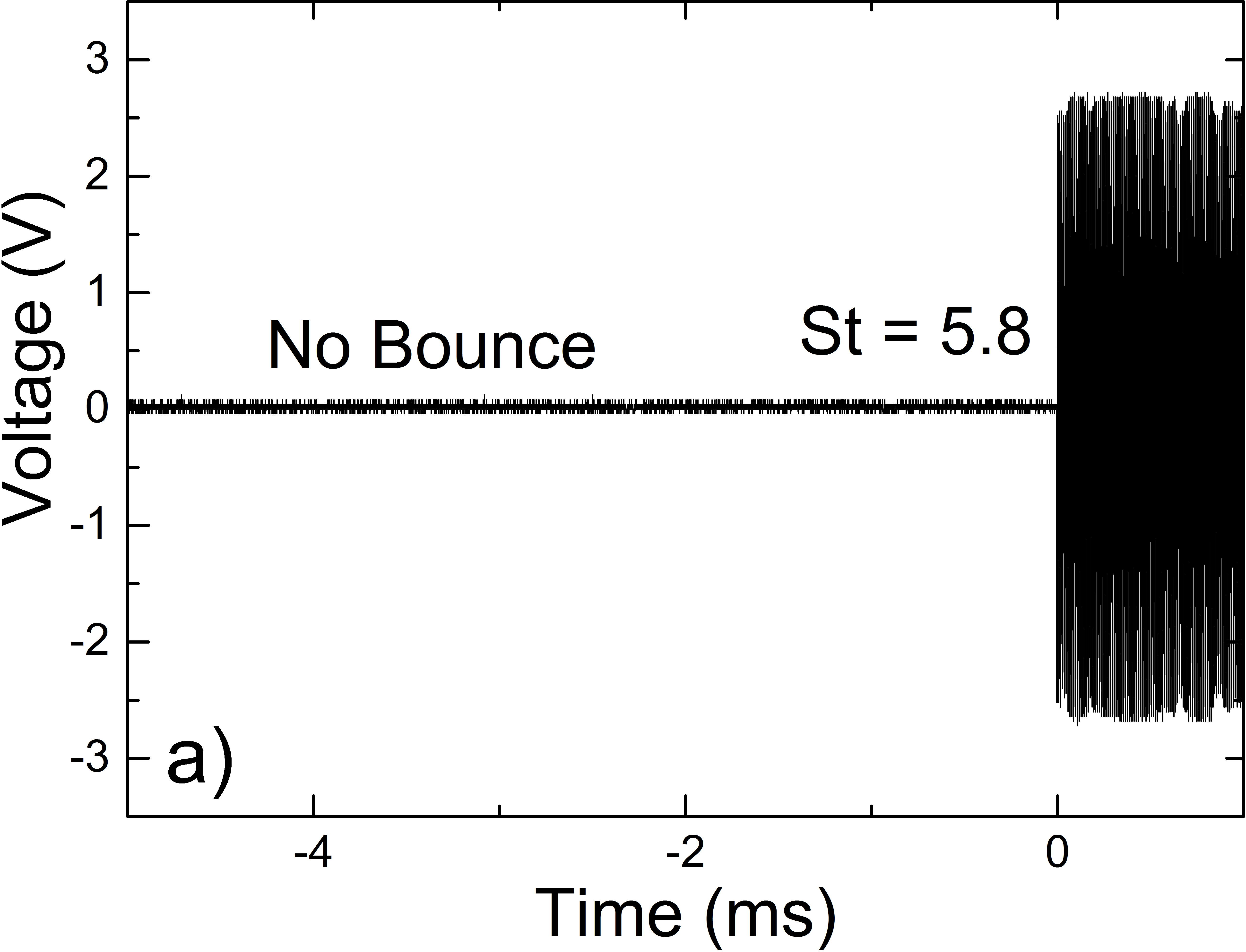}\hspace{1cm}
\includegraphics[width=6.525cm,height=5.25cm]{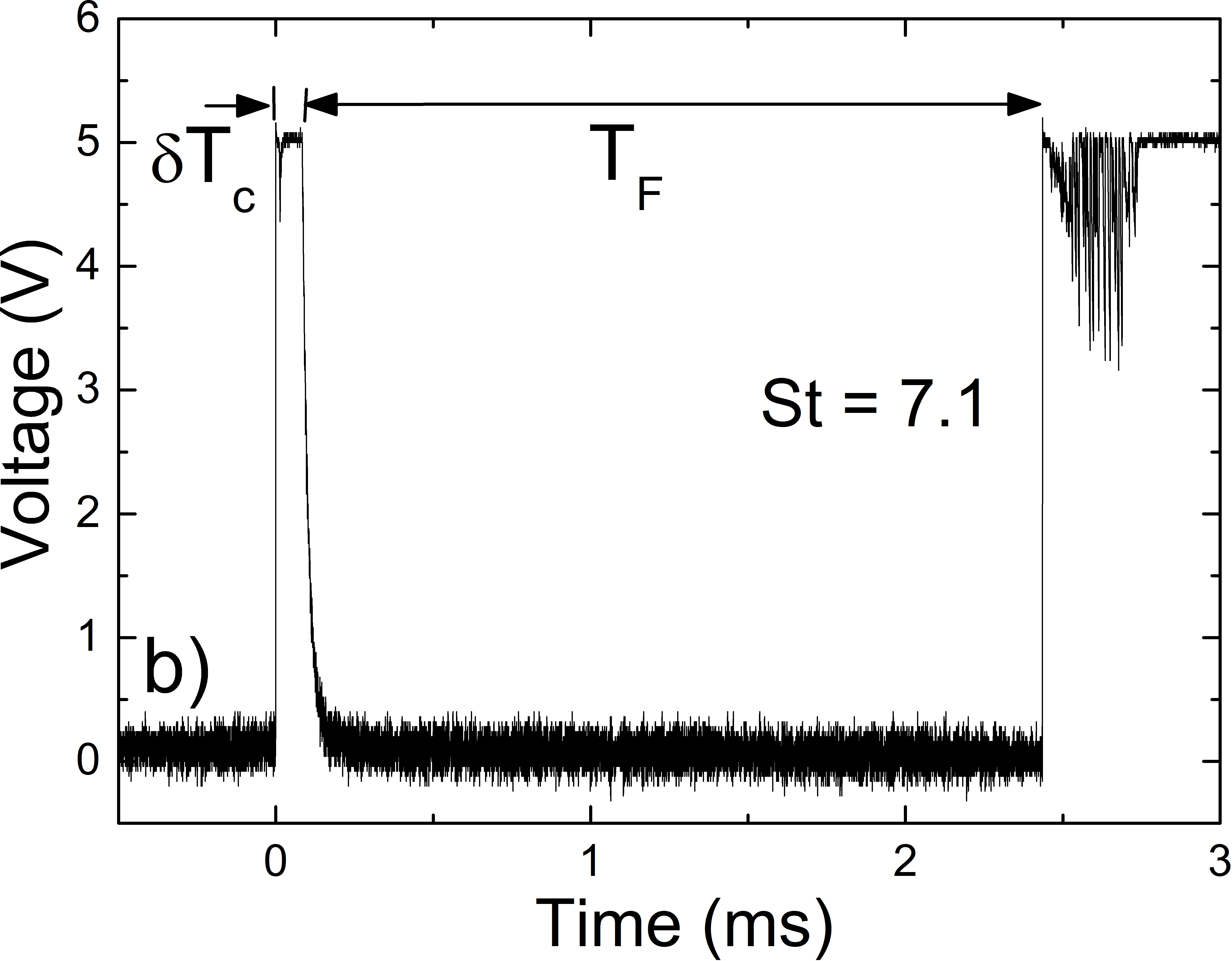}\vspace{0.5cm}

\includegraphics[width=6.3cm,height=5.25cm]{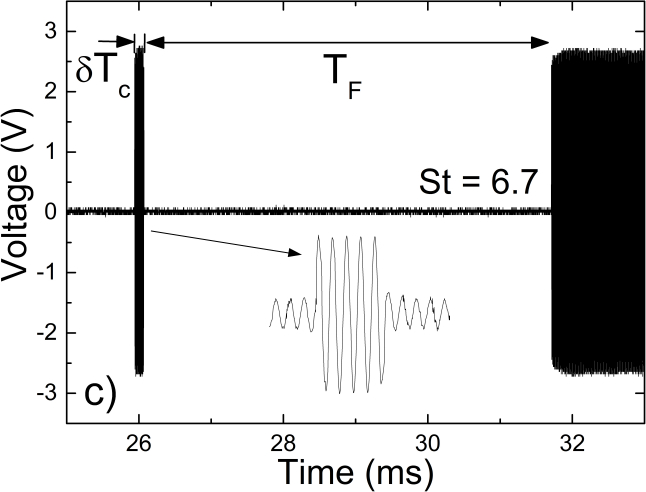}\hspace{1cm}
\includegraphics[width=6.9cm,height=5.25cm]{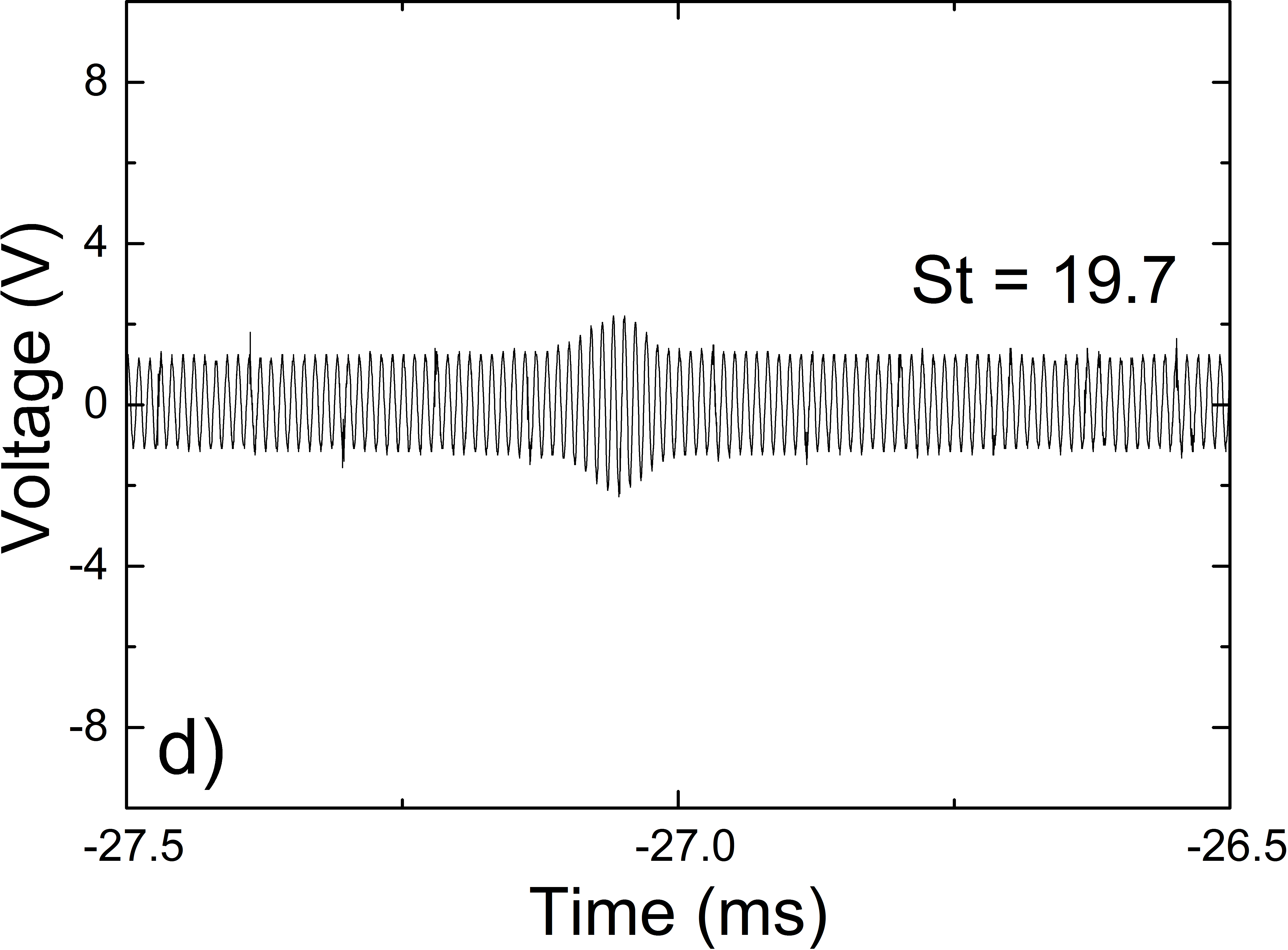}

\caption{\label{fig:contact}Voltage versus time graphs for increasing $St$ with AC and DC voltages. (a) At $St=5.8$ with an AC applied voltage, no bounce occurs, but electrical contact is made in a finite time. This is the generic behaviour for $St<St_c$. (b) At $St=7.1$ and (c) at $St=6.7$ with DC and AC voltages, respectively, we see a collision, and then a period of no contact, followed by permanent contact. The contact has much lower resistance than the series resistance i.e. $R_c\ll R_s$. This is the generic behaviour for  $St>St_c$. For $St>St_c$, we also see the case shown in (d) where there is a high resistance contact ($R_c$ is of the same order as $R_s$) (more details are in appendix B).}
\end{center}
\end{figure}

In Figure~\ref{fig:contact} we show examples of the collision obtained by applying both AC (Figure~\ref{fig:contact}a,~\ref{fig:contact}c and~\ref{fig:contact}d) and DC voltages  (Figure~\ref{fig:contact}b) between plate and sphere. At very low Stokes number (panel a, $St=5.8$), we observe no bounces. The first electrical contact persists for all time. At larger values of $St$, as shown in Figure~\ref{fig:contact}b and~\ref{fig:contact}c, the ball makes metallic contact for a finite contact time, $\delta T_c$. It then breaks contact and is in the fluid for a flight time $T_F$ before settling into permanent electrical contact. There is a clear separation of scale between the contact time, $\delta T_c$ (tens of $\mu$-sec) and the  flight time, $T_F$ (tens of milli-sec). Contact can sometimes be noisy (as seen at late times in Figure~\ref{fig:contact}b, presumably due to rolling or rocking of the sphere. Such differential motions have been studied for a sphere in a rotating cylinder \cite{ashmore2005cavitation,yang2006motion}. Finally, we see instances, as shown in Figure~\ref{fig:contact}d, where the contact resistance ($R_c$) between the ball and the plate is comparable to the external series resistance ($R_s$).

To test if electrical forces play a role in contact mechanics, we compared
our results with high frequency AC (100 to 500 kHz) to results with a DC voltage, and made measurements as a function of the amplitude of the applied voltage. We chose silicone oil as the working fluid due to its high dielectric breakdown voltage ($>40$ MV/m). There appears to be no systematic effect of the amplitude or frequency of the AC voltage or the choice of AC or DC voltage on the occurrence of contact, the duration of contact and on the intervals between bounces at a given $St$ (as detailed in appendix B). We thus conclude that our results are not due to dielectric breakdown. All observed contacts are resistive, that is, with no phase shift between the applied and measured voltages. However, we do point out that the relative frequency of low versus high-resistance contacts can be affected by many factors such as the impurity content of the oil, ageing of the surface electrical properties, and the topography  of the sphere following repeated collisions. To achieve consistent results, we report data taken under a set of fixed conditions. 

\begin{figure}
\begin{center}
\includegraphics[width=11.0cm,height=9.0cm]{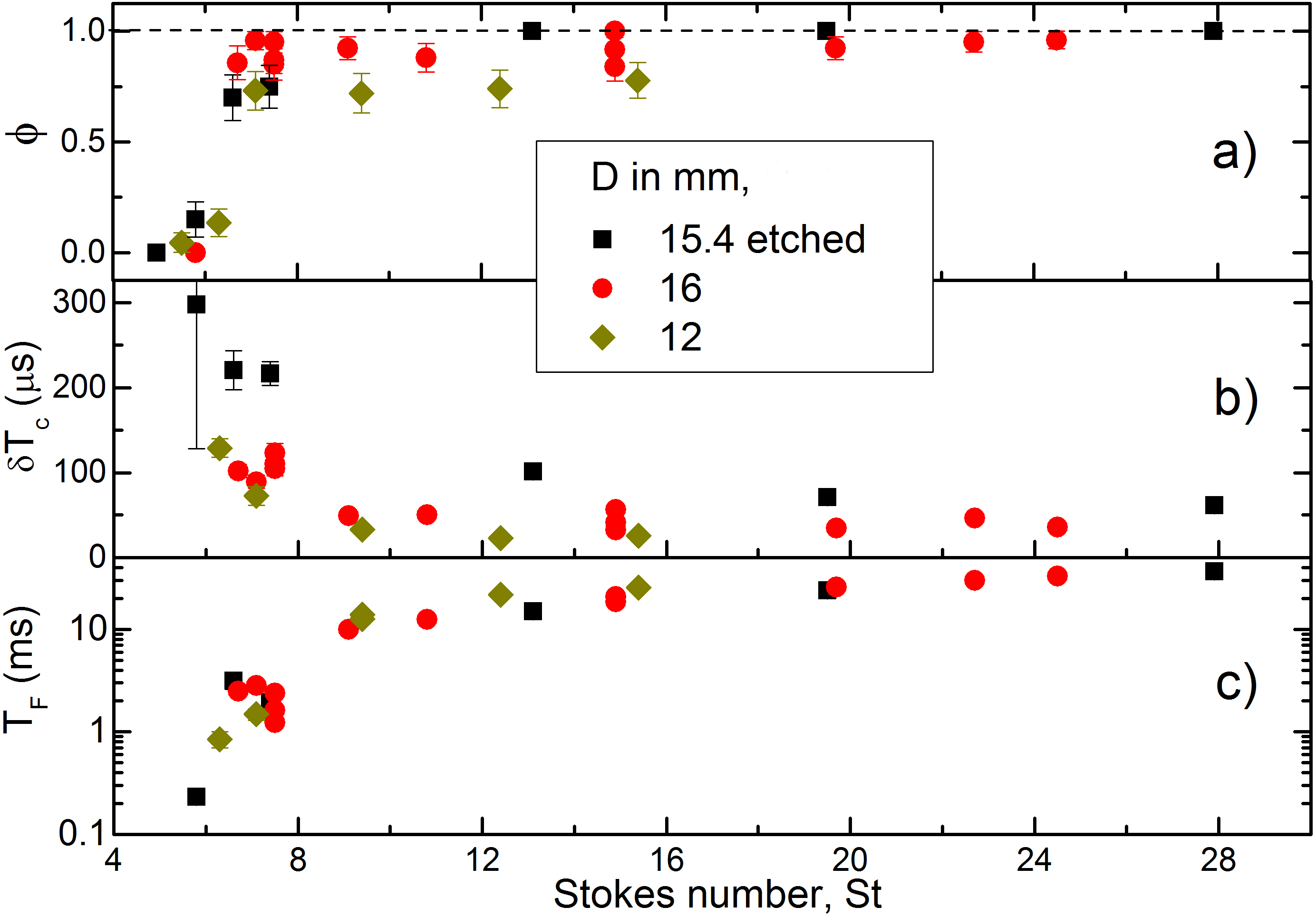}
\caption{\label{fig:AvgContact}Contact fraction ($\phi$), contact time ($\delta T_c$), and flight time ($T_F$) vs Stokes number for the 15.4 mm etched ball, 16 mm and 12 mm unetched balls. Here we report data only for low resistance contacts ($R_c \ll R_s$)  as in Figure 3b and 3c, however the trends are unaffected if we include more resistive contacts (see appendix B) as in Figure 3d. Error bars are the standard deviation of measurements in panel a) and standard error of measurements in panel b) and c) taken at fixed $St$. Wherever the error bars are not visible, errors are smaller than symbol size.}
\end{center}
\end{figure}

The major qualitative result in Figure~\ref{fig:contact} is that the ball makes direct mechanical contact with the plate during the bounce, in contrast to expectations based on elastohydrodynamic theory \cite{davis1986elastohydrodynamic}. Next, we explore the nature of that contact. 

We define contact fraction, $\phi$, as the fraction of experiments at a particular Stokes number in which the sphere made a low resistance contact with the plate's surface during the bounce, i.e., collisions as shown in Figure~\ref{fig:contact}b and~\ref{fig:contact}c. The value of $\phi$ rises sharply from zero above a critical Stokes number $St_c$, which is the same for all three types of spheres we used. Thus solid-on-solid contact occurs even just above the threshold of bouncing. The value of $St_c \approx 6.2 \pm 0.5$ that marks the transition to bouncing with mechanical contact is consistent with the bouncing transition observed in previous experiments \cite{gondret1999experiments,gondret2002bouncing,joseph2001particle}. We refer to $\phi$ as the contact fraction, even though it is actually a lower bound on the contact fraction, in that high resistance events are not included. Including high resistance events, as shown in Figure~\ref{fig:voltage}, Appendix B, does not change any qualitative trends.

In Figure~\ref{fig:AvgContact}b we show data for the duration of contact, $\delta T_c$. The contact time $\delta T_c$ decreases as the Stokes number is increased above $St_c$. The relatively small change of $\delta T_c$ is consistent with calculations for a Hertzian, elastic impact \cite{landau1986theory} which predict a very weak dependency of contact time on velocity, $\delta T_c$ $\propto$ $(U R_\textit{eff})^{-1/5}$, where $R_\textit{eff}$ is the effective radius at the point of contact. For perfectly smooth spheres, $R_\textit{eff}=R$, whereas $R_\textit{eff}$ will be smaller when a bump on the sphere is presented to the plane. Contact times for the etched spheres are slightly longer than those for the unetched spheres (which have very similar contact times for both sizes of sphere). Roughness at the point of contact, rather than the sphere radius, possibly sets the relevant curvature at impact and influences the contact time (see appendix C).
In Figure~\ref{fig:AvgContact}c we show the duration between the bounce and the next collision, which we refer to as the flight time $T_F$. This is a measure of the kinetic energy with which the ball rebounds from the plate. As expected, this is an increasing function of ($St-St_{c}$). Furthermore, there is little or no variation with the type of sphere used, as opposed to the data for $\delta T_c$. This implies that the details of the solid contact may not affect the total dissipation, as elaborated below.

\begin{figure}
\begin{center}
\includegraphics[width=11.0cm,height=9.0cm]{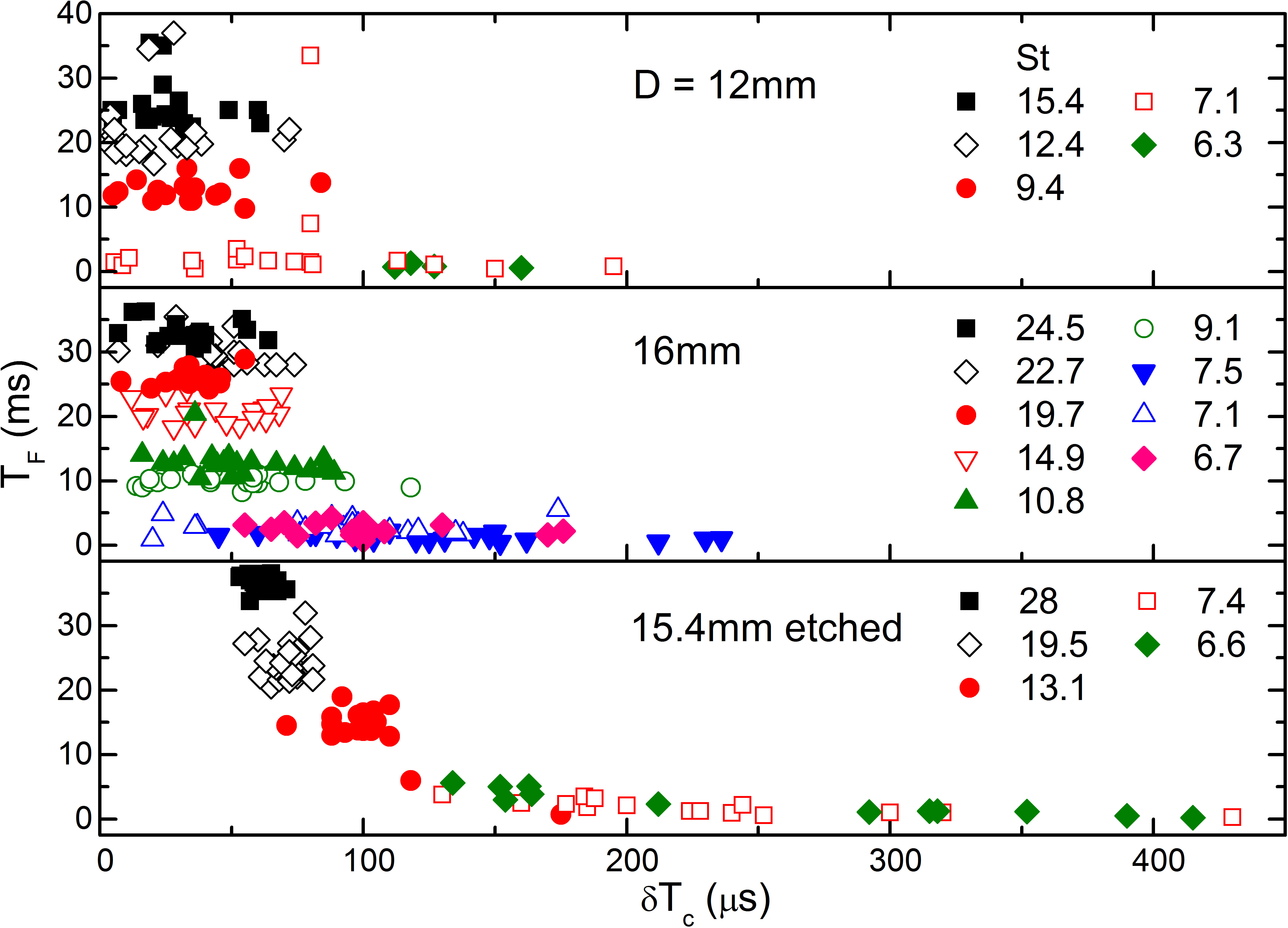}
\caption{\label{fig:TimeDistributions}Flight time vs contact time for $D=16$ mm un-etched, 12 mm and 15.4 mm etched balls. Notice the increase in the scatter at lower $St$.}
\end{center}
\end{figure}

Thus far, we have discussed mean values of the contact duration and flight times, averaged over experimental trials at fixed $St$. However, the distribution of $\delta T_c$ and $T_F$ reveals the source of dissipation. In Figure~\ref{fig:TimeDistributions} we plot $T_F$ versus $\delta T_c$ for $St$ varying from just above $St_c$ to about $St=28$. In most cases, we find a  broad distribution of contact times varying by up to an order of magnitude, at a fixed value of $St$. Presumably this reflects the variation in the local topography of the sphere. We however find much smaller variability in the flight time, particularly at larger $St$. Thus the total dissipation in the sphere-wall encounter, as reflected by the flight time, is not strongly affected by the duration of solid-on-solid contact. This indicates that despite solid contact, the bulk of kinetic energy is lost to fluid dissipation. 

At the largest Stokes numbers a different trend sets in, most clearly observed in the etched spheres: the contact time $T_c$ becomes narrowly distributed. We suggest that at large impact speeds, the effective radius of the Hertzian contact increases, and averages over the roughness of the etched sphere (see appendix C). Thus collisions at different locations become similar in their contact mechanics and no longer depend on local roughness. We anticipate another regime -- as in a ball bouncing in air -- at even higher $St$ where the dissipation becomes solid-dominated, but we do not observe that regime. At the largest $St$, we observe plastic deformation in the form of pitting at the point of impact (more details are in appendix C).

While the data make a clear case for solid-on-solid contact in all the spheres used, the $St-$dependence of the details of the bounce appear to be influenced by the surface quality of the sphere. This leads one to question whether our conclusions are valid for spheres that are even smoother than the ball bearings employed here. The following calculation is illuminating in this regard.

\section{Lubrication approximation}

\begin{figure}
\begin{center}
\includegraphics[width=11.0cm,height=8.5cm]{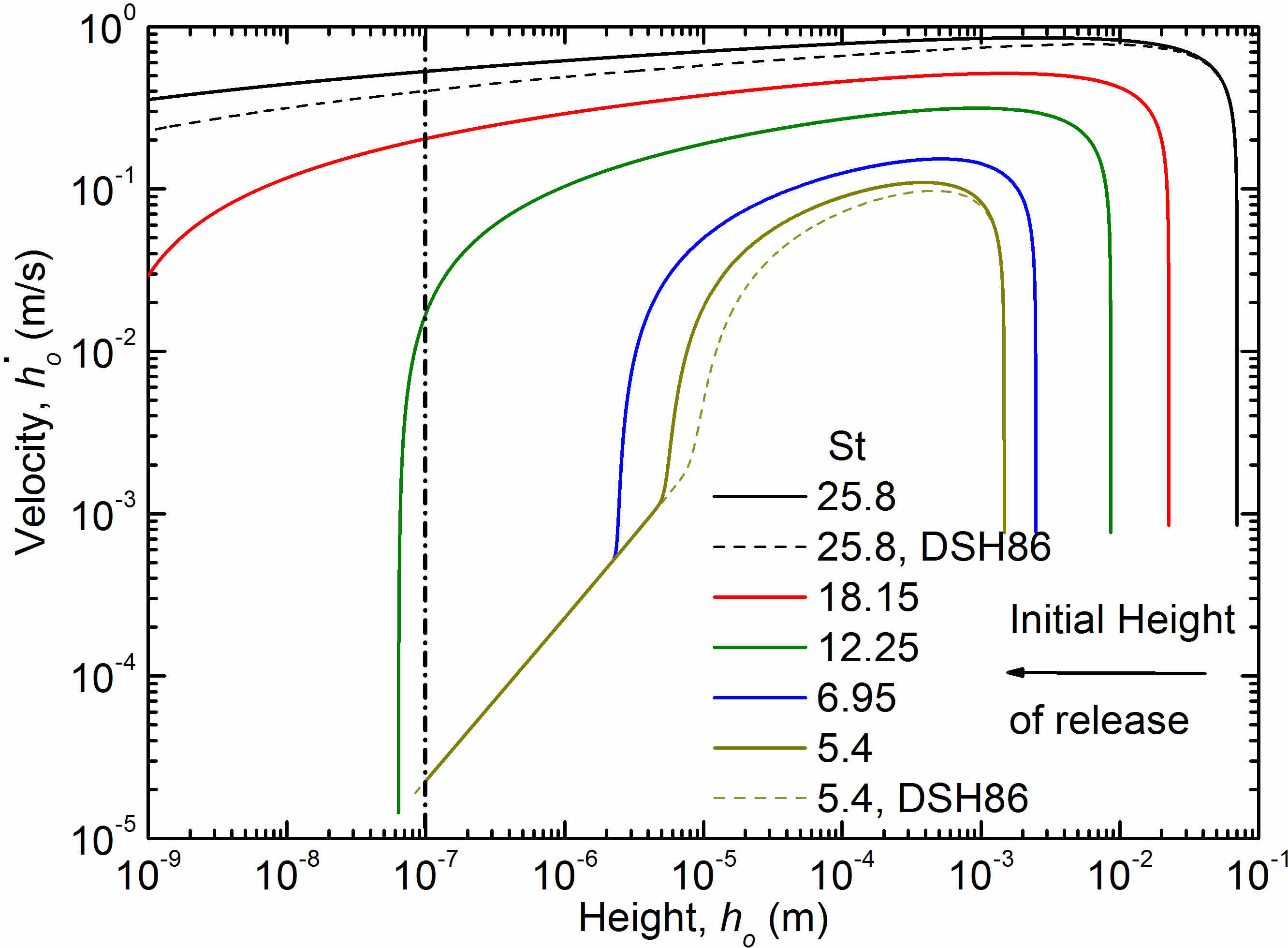}
\caption{\label{fig:Lubrication}Velocity ($\dot h$) versus height ($h_o$) of a rigid smooth sphere approaching a rigid plane calculated from equation of motion \ref{vertical_balance}. The results shown are for a steel sphere of diameter 15.4 mm, dropped from different heights yielding the labelled nominal values of $St$. The solid curves are obtained by using Equation \ref{horizontal_balance} and Equation \ref{spheroid} to get the net pressure force, and then numerically solving Equation \ref{vertical_balance}. The curves for various Stokes numbers cross a given height with different velocities, a typical example is shown by the vertical dash-dot line at $h_o = 100$ nm. At this height, the velocity for the curve labelled $St=12.25$ is three orders of magnitude higher than those for the curves shown for lower $St$. This implies that even for extremely smooth surfaces, a physical contact will occur prior to bouncing for a high enough Stokes number.
The dashed curves at $St=25.8$ and $St=5.4$ are obtained by approximating the sphere as a paraboloid near $r=0$, to compare with previous calculations in DSH86: \cite{davis1986elastohydrodynamic} that use this approximation.}
\end{center}
\end{figure}

We write the equation of motion for a smooth, rigid sphere approaching a plane under the lubrication approximation \cite{davis1986elastohydrodynamic}. The vertical distance between ball and plate, $h_o=h(0,t)$, varies as
\begin{equation}
\rho_s V \ddot h_o = (\rho_s - \rho_f)Vg - 3\pi \mu D \dot h_o (1 + 0.15 Re^{0.687}) - F_p,
\label{vertical_balance}
\end{equation}
where $Re$ is defined by the instantaneous velocity of the sphere, $\dot h_o$ (see Figure~\ref{fig:setup}a). $F_p$ denotes the upward pressure force exerted by the fluid layer. As in Equation 2, we use an empirical formula for the viscous drag in the absence of a bottom plate. The horizontal force balance between pressure gradient and viscous forces is given by:
\begin{equation}
\frac{\partial p}{\partial r} + \mu \frac{\partial^2 u_r}{\partial z^2} = 0,
\label{horizontal_balance}
\end{equation}
where $p(r,t)$ is the pressure. The radial velocity of the fluid, $u_r$, squeezing out between the plate and the sphere is assumed to have a parabolic profile \cite{leal2007advanced}. To obtain the net pressure force $F_p$, we solve Equation \ref{horizontal_balance} using the relation for surface profile of a sphere:
\begin{equation}
 h(r,t) = h_o(t) + R - \sqrt{R^2 -r^2},
 \label{spheroid}
\end{equation}
where $r=R$ is the the radius of the sphere. Equation \ref{vertical_balance} is numerically solved to obtain $\dot{h_o}$ as a function of $h_o$, and is shown by solid curves in Figure~\ref{fig:Lubrication}. In the calculation, we follow previous work by Davis \cite{davis1986elastohydrodynamic} except that they had made the further assumption that the region near $r=0$ may be modelled as a paraboloid rather than a sphere. The approximation of a paraboloid allows for analytic solutions, as shown by the dashed curves in Figure~\ref{fig:Lubrication}, that are quantitatively close to our numerical solutions for a sphere.

Even within these approximate treatments, where $p$ diverges as $h_o\rightarrow0$, the velocity of approach is significant at a roughness cut-off scale as small as 1 nm, for large enough $St$. Thus the calculation indicates that contact occurs even when the spheres are atomically smooth.

The lubrication approximation ignores vertical flows and radial gradients. However, for this geometry, near contact, both these assumptions can be violated as the radial velocity vanishes and radial gradients can be large. As this is the region closest to the bottom suface, further quantitative comparison with experiments will require going beyond elastohydrodynamic lubrication theory. However, direct solid contact is likely to be a general feature, even close to the threshold of bouncing. In this regime, the dissipation remains dominated by fluid mechanics, while the duration of the contact is largely controlled by surface and bulk properties of the solids. Even though solid dissipation does not play a prominent role, the presence of solid-to-solid contact is of great significance in contexts such as wear, charge transfer, or chemical reactivity of solids in suspension. 

\section{Acknowledgement}

We acknowledge funding from TCIS Hyderabad, ICTS Bangalore, the APS-IUSSTF for a travel grant to SKB, and NSF-DMR 120778 and NSF-DMR 1506750 for support at UMass Amherst. We are grateful to John Nicholson and the Keck Nanotechnology facility for profilometry and to Bharath B and G. U. Kulkarni from JNCASR for optical profilometry. 

\section{Appendix A: Roughness}

\begin{figure}
\begin{center}
\includegraphics[width=11.0cm,height=8.5cm]{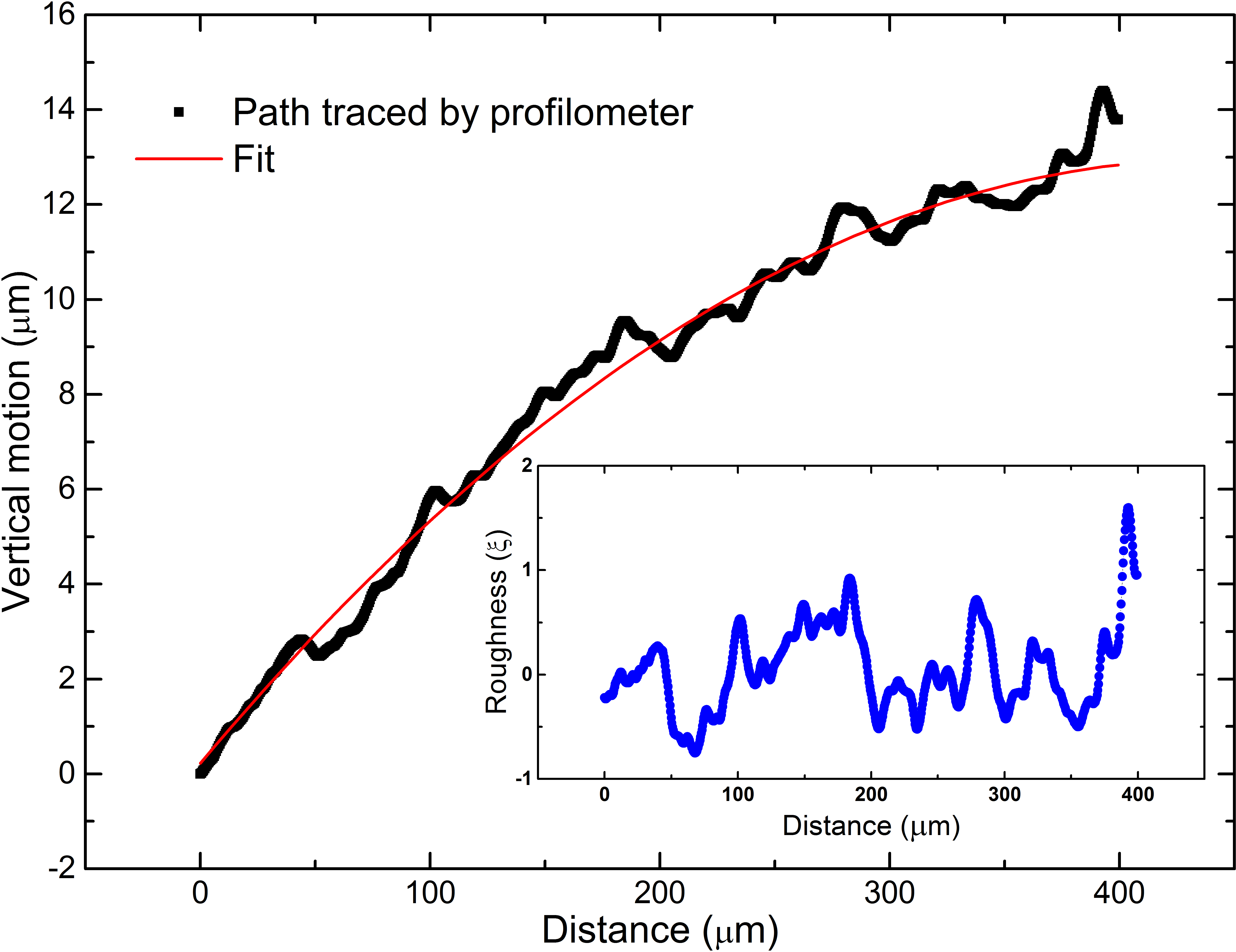}
\caption{\label{fig:fit} A fit for the 15.4 mm etched sphere on the path traced by the profilometer as a function of position along a line on the sphere's surface is shown in red. The fit gives us the  global curvature of the path and the local roughness of the sphere is obtained by subtracting the surface profile from this fit. The roughness ($\xi$) is shown in blue in the inset.}
\end{center}
\end{figure}

The roughness profile of the spheres used was measured with a diamond-tip Dektak contact profilometer with lateral resolution of 0.5 $\mu$m and vertical range of 65.5 $\mu$m. The vertical resolution of the profilometer is $\approx 1$ nm. The profilometer tip travels along a path of length $\sim 0.4$ mm along the surface of the sphere. We fit a circle to the height along the path traced by profilometer. This yields the global radius of  curvature  along this path as shown in Figure~\ref{fig:fit}. The deviations from this fit give the local roughness, $\xi$ of the topography of the surfaces. As can be seen from the statistics of $\xi$ given in Table 1, the etched 15.4 mm sphere's surface was found to be most rough; as discussed in the experiment section. It is also the least heterogeneous, with relatively uniform roughness across the surface. 

The plate roughness was measured using optical profilometry. The rms roughness of the plate was found to be around 0.29 $\mu$m and the average of the maxima of peak to valley height from 12 random samples of the plate was observed to be 3.2 $\mu$m

\begin{center}
\begin{tabular}{| c | c | c | c |}
\multicolumn{4}{c}{Table 1. Roughness values} \vspace{0.3cm}\\\hline

  D = & 12 mm & 16 mm & 15.4 mm \vspace{0.0cm}\\\hline
 RMS ($\mu$m) & 0.25 & 0.22 & 0.5 \vspace{0.0cm}\\\hline 
 Max positive deviation ($\mu$m) & 1.1 & 0.68 & 1.66 \vspace{0.0cm}\\\hline
 Max negative deviation ($\mu$m) & -1.78 & -1.77 & -1.4\vspace{0.0cm}\\\hline
\end{tabular}
\end{center}

\section{Appendix B: Effect of electrical forces}

In order to verify that electrical forces do not play any role in the mechanics of contact, we compared our results for a fixed Stokes number $St$ at different AC voltages and frequency. We did not find any systematic dependence in duration of contact or flight time with applied voltage or frequency. Results obtained with DC measurements were found to be similar to AC measurements.

\begin{figure}
\begin{center}
\includegraphics[width=14.6cm,height=12.0cm]{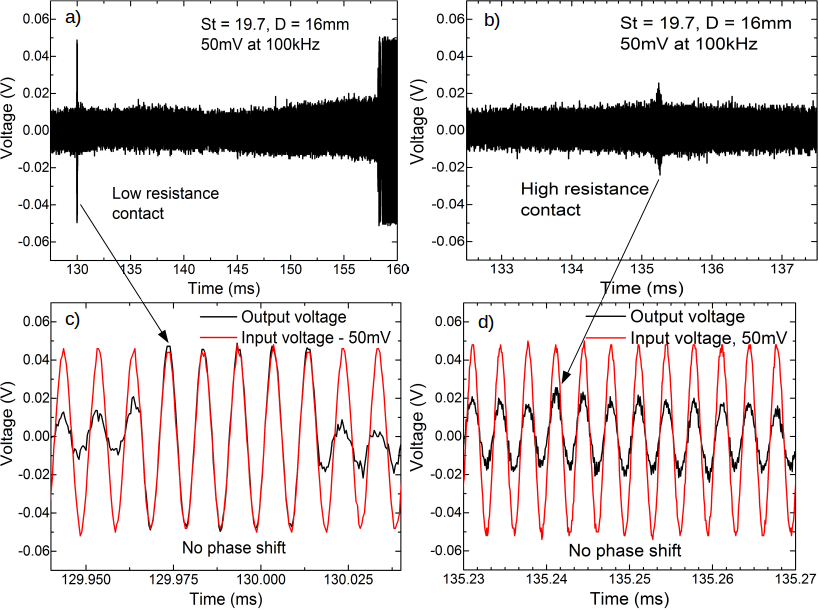}

\caption{\label{fig:phase}a) Low-resistance contact and b) high-resistance contact at 50 mV, 100 kHz and $St=19.7$. Panel c) and d) are the zoomed-in version of contacts shown in panel a) and b). The voltage shown in black is measured by a digital oscilloscope across a 220 k$\Omega$ resistor. In panel a) and c), the output voltage is equal to the source voltage (shown in red) of 50 mV suggesting that there is no voltage drop across the contact and the contact has very low resistance. However in panel b) and d), the output voltage is lower than the source voltage suggesting that some contacts obtained during the experiments have high resistance. Notice the zero phase shift in both cases.}
\end{center}
\end{figure}

We used the same circuit as discussed before, Figure~\ref{fig:setup}a. The output voltage was measured across a 220 k$\Omega$ resistor, $R_s$. This meant that if the contact resistance, $R_c$, is very low all the voltage drop would occur across $R_s$. This indeed was observed in a majority of the  experiments. An example has been shown in Figure~\ref{fig:phase}a and a zoomed-in version of this contact is visible in Figure~\ref{fig:phase}c. The output voltage (shown in black) is equal to applied voltage of 50 mV (shown in red) suggesting that no voltage drop occurs across contact. Such contacts will be called ``low resistance contacts". In the rest of the experiments we observed a significant voltage drop occurring across the contact. Hence, the voltage measured across the oscilloscope is less than the applied voltage as is visible in Figure~\ref{fig:phase}b and~\ref{fig:phase}d. This suggests that $R_c$ is of the same order as $R_s$ in these cases. Such contacts were termed as ``high resistance contacts". The measurements shown in Figure~\ref{fig:phase} were taken at $St = 19.7$, 100 kHz for 16 mm sphere.
Contacts were found to be purely resistive as can be seen from the fact that there is zero phase difference between the applied voltage and output voltage across $R_s$ in Figure~\ref{fig:phase}, panels c and d.

\begin{figure}
\begin{center}
\includegraphics[width=11.0cm,height=9.0cm]{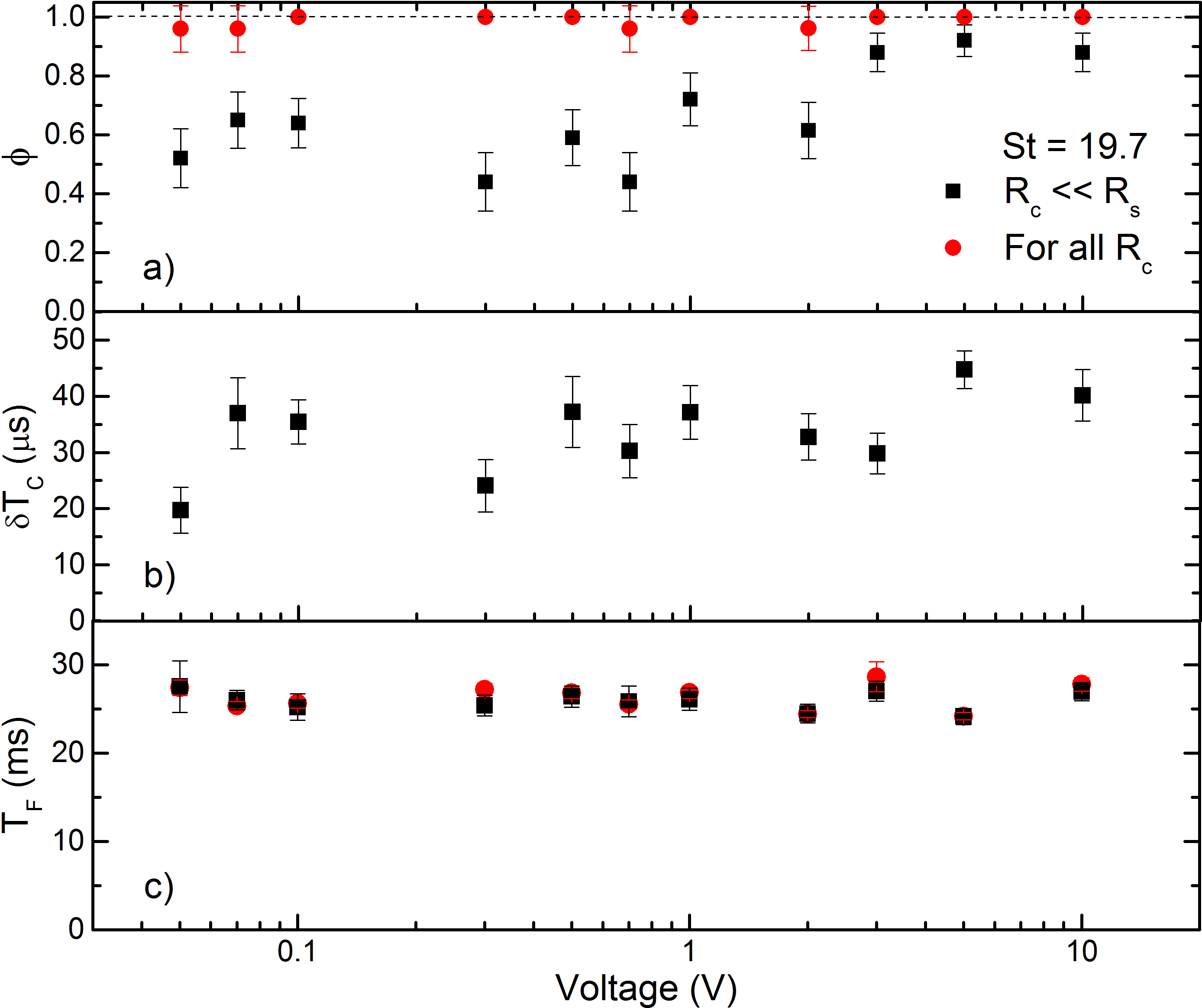}
\caption{\label{fig:voltage}There appears to be no systematic effect of the electrical voltage on the duration of contact (panel b) or on the intervals between bounces (panel c). This experiment was carried out at 300 kHz, $St = 19.7$. The total measured contact fraction (panel a) remains close to one. The fraction of low resistance contacts (black squares) is lower. Error bars are the standard deviation of measurements in panel a) and standard error of measurements in panel b) and c) taken at fixed Stokes numbers. Wherever the error bars are not visible, errors are smaller than symbol size.}
\end{center}
\end{figure}

In Figure~\ref{fig:voltage}, we have made measurements as a function of the amplitude of the applied AC voltage for same Stokes number as above ($St=19.7$) and $D = 16$ mm. The applied AC voltage was varied from 50 mV to 10 V. The fraction of low resistance contact ($R_c << R_s$) was found to be lower than unity, implying that high resistance contact occurs in some of the trials. When high resistance contacts are also included, the contact percentage becomes one and does not vary with applied voltage. We point out that the fraction of high resistance contacts is variable and can be affected by many factors such as the impurity content of the oil, and ageing of the surface electrical properties and topography of the sphere following repeated collisions.

\begin{figure}
\begin{center}
\includegraphics[width=11cm,height=8.5cm]{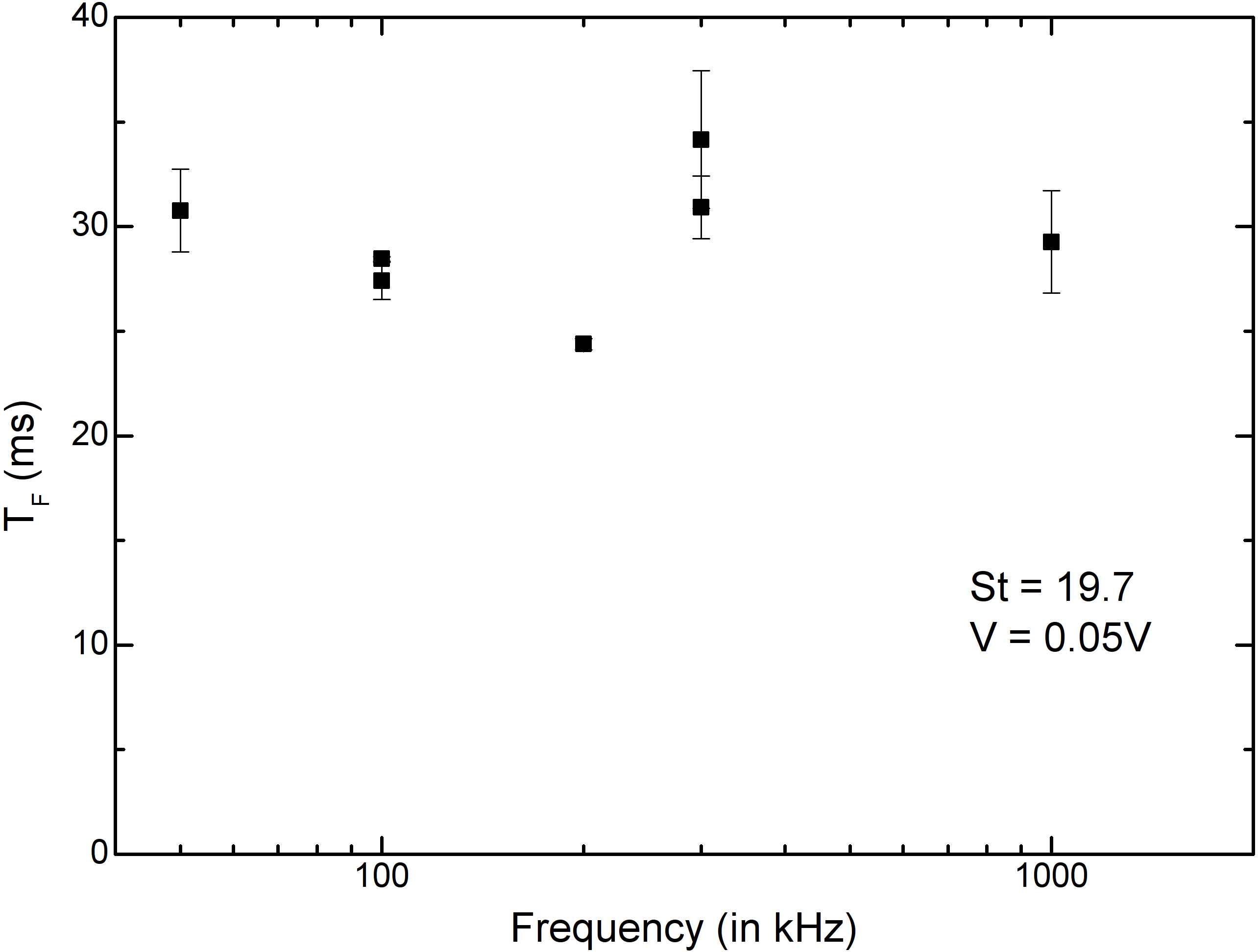}
\caption{\label{fig:frequency}Variation in flight time with frequency. No systematic trend was observed and contact fraction was found to be equal to one for all values of frequencies.}
\end{center}
\end{figure}

As is evident from Figure~\ref{fig:voltage}, no systematic dependence of contact time (panel b) and flight time (panel c) on electrical voltage was observed. This suggests that electrical forces do not play a role in the dynamics of the collision. In particular, there is no observed change in the nature of the electrical signal, arguing that dielectric breakdown of silicone oil does not occur during the collision process. The literature value of the breakdown voltage is high but finite ($>40$ MV/m), so perhaps breakdown is avoided due to the very short time-scales of interaction. We thus infer that the values of contact time, flight time and contact fraction reported in this article are independent of voltage applied.

The dependence on frequency (50 to 1000 kHz) of applied voltage was also studied and is shown in Figure~\ref{fig:frequency}. No systematic trend in duration of contact, flight time and contact fraction was observed. Figure~\ref{fig:frequency} shows that there is no significant change in flight time with frequency. As mentioned before, the contact fraction, when both low resistance contacts and high resistance contacts are included, was found to be close to one for all values of applied frequencies at $St>St_c$.

\section{Appendix C: Hertzian Calculation}

For a perfect normal collision between two smooth, perfectly elastic spheres of radii $R_i$, masses $m_i$, Young's modulus $E_i$, and Poisson ratio $\sigma_i$ (i=1,2), the closest approach of the spheres during mechanical contact can be calculated by Hertzian theory \cite{landau1986theory} and is given by 

\begin{figure}
\begin{center}
\includegraphics[width=8.0cm,height=7.0cm]{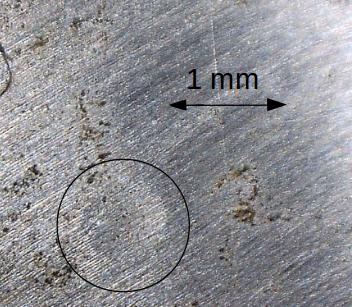}
\caption{\label{fig:dent}Craters visible inside the marked circle. The sizes of the crater on the plate is around 700 $\mu$m for $St = 28$.}
\end{center}
\end{figure}

\begin{equation}
    b = \Big(\frac{m_\textit{eff}}{k}\Big)^{2/5} v^{4/5},
\end{equation}
where b is the maximum overlap of the spheres ($b$ = distance between the centres of the sphere subtracted from $(R_1 + R_2)$) and $v$ is the relative velocity between the spheres just before contact. Here $m_\textit{eff}$ is the reduced mass of the spheres and is given by
\begin{equation}
    \frac{1}{m_\textit{eff}} = \frac{1}{m_1} + \frac{1}{m_2},
\end{equation}
 $k$ is the effective modulus of the spheres, given by
\begin{equation}
    k = \frac{4}{5 D_0} \sqrt\frac{R_1 R_2}{R_1 + R_2},
\end{equation}
where
\begin{equation}
  D_0 = \frac{3}{4} \Big(\frac{1-\sigma_1 ^2}{E_1} + \frac{1-\sigma_2^2}{E_2}\Big).
\end{equation}
In our set-up, one of the spheres has been replaced by a plate, so $1/R_2 = 0$ and $1/m_2 = 0$.

For a smooth sphere of diameter $15.4$ mm, the depth of the contact formed due to collision of the sphere with the plate for an impact velocity corresponding to $St=28$ is $\approx 13.5~ \mu$m. The radius of the contact area is then given by
\begin{equation}
  a = \sqrt{2Rb}.
\end{equation}
This gives the diameter of the contact area $= 2a \approx 900~ \mu$m. Of course, due to lubrication forces, the actual impact velocity will be much lower, so these calculations considerably over predict the size of the crater made. 

A different possibility is that instead of the radius of the sphere setting the geometry of the collision, the local radius of curvature set by the roughness of the sphere is the relevant radius. If the characteristic height of bumps is $q$ and the characteristic lateral size is $l$, the local radius of curvature will be given by
\begin{equation}
  R_\textit{eff} = \frac{1}{2q} \Big(\frac{l}{2}\Big)^2.
\end{equation}
For $q=1~\mu$m and $l=100~\mu$m, $R_\textit{eff} = 900~\mu$m. If we use this radius in Hertzian calculation for collision, we get the depth and radius of the contact area to be $\approx 20~\mu$m and $200~\mu$m, respectively. 

At the largest Stokes numbers, we see that the sphere's impact on the plate leaves small craters (Figure~\ref{fig:dent}) $\approx 700~\mu$m, thus the Hertzian calculation must be replaced by a calculation involving material plasticity. Finally, we recall that all these estimates have used the Stokes number computed without wall effects, thus the actual impact velocity will be lower. These estimates therefore overpredict the size of the region of contact.

\bibliography{Main_paper}

\begin{thebibliography}{14}%
\makeatletter
\providecommand \@ifxundefined [1]{%
 \@ifx{#1\undefined}
}%
\providecommand \@ifnum [1]{%
 \ifnum #1\expandafter \@firstoftwo
 \else \expandafter \@secondoftwo
 \fi
}%
\providecommand \@ifx [1]{%
 \ifx #1\expandafter \@firstoftwo
 \else \expandafter \@secondoftwo
 \fi
}%
\providecommand \natexlab [1]{#1}%
\providecommand \enquote  [1]{#1}%
\providecommand \bibnamefont  [1]{#1}%
\providecommand \bibfnamefont [1]{#1}%
\providecommand \citenamefont [1]{#1}%
\providecommand \href@noop [0]{\@secondoftwo}%
\providecommand \href [0]{\begingroup \@sanitize@url \@href}%
\providecommand \@href[1]{\@@startlink{#1}\@@href}%
\providecommand \@@href[1]{\endgroup#1\@@endlink}%
\providecommand \@sanitize@url [0]{\catcode `\\12\catcode `\$12\catcode
  `\&12\catcode `\#12\catcode `\^12\catcode `\_12\catcode `\%12\relax}%
\providecommand \@@startlink[1]{}%
\providecommand \@@endlink[0]{}%
\providecommand \url  [0]{\begingroup\@sanitize@url \@url }%
\providecommand \@url [1]{\endgroup\@href {#1}{\urlprefix }}%
\providecommand \urlprefix  [0]{URL }%
\providecommand \Eprint [0]{\href }%
\providecommand \doibase [0]{http://dx.doi.org/}%
\providecommand \selectlanguage [0]{\@gobble}%
\providecommand \bibinfo  [0]{\@secondoftwo}%
\providecommand \bibfield  [0]{\@secondoftwo}%
\providecommand \translation [1]{[#1]}%
\providecommand \BibitemOpen [0]{}%
\providecommand \bibitemStop [0]{}%
\providecommand \bibitemNoStop [0]{.\EOS\space}%
\providecommand \EOS [0]{\spacefactor3000\relax}%
\providecommand \BibitemShut  [1]{\csname bibitem#1\endcsname}%
\let\auto@bib@innerbib\@empty
\bibitem [{\citenamefont {Reynolds}(1886)}]{reynolds1886theory}%
  \BibitemOpen
  \bibfield  {author} {\bibinfo {author} {\bibfnamefont {O.}~\bibnamefont
  {Reynolds}},\ }\bibfield  {title} {\enquote {\bibinfo {title} {On the theory
  of lubrication and its application to Mr. Beauchamp Tower's experiments,
  including an experimental determination of the viscosity of olive oil},}\
  }\href@noop {} {\bibfield  {journal} {\bibinfo  {journal} {Proc. R. Soc.
  Lond.}\ }\textbf {\bibinfo {volume} {40}},\ \bibinfo {pages} {191} (\bibinfo
  {year} {1886})}\BibitemShut {NoStop}%
\bibitem [{\citenamefont {Davis}\ \emph {et~al.}(1986)\citenamefont {Davis},
  \citenamefont {Serayssol},\ and\ \citenamefont
  {Hinch}}]{davis1986elastohydrodynamic}%
  \BibitemOpen
  \bibfield  {author} {\bibinfo {author} {\bibfnamefont {R.~H.}\ \bibnamefont
  {Davis}}, \bibinfo {author} {\bibfnamefont {J.-M.}\ \bibnamefont
  {Serayssol}}, \ and\ \bibinfo {author} {\bibfnamefont {E.~J.}\ \bibnamefont
  {Hinch}},\ }\bibfield  {title} {\enquote {\bibinfo {title} {The
  elastohydrodynamic collision of two spheres},}\ }\href@noop {} {\bibfield
  {journal} {\bibinfo  {journal} {J. Fluid Mech.}\ }\textbf {\bibinfo {volume}
  {163}},\ \bibinfo {pages} {479} (\bibinfo {year} {1986})}\BibitemShut
  {NoStop}%
\bibitem [{\citenamefont {Gondret}\ \emph {et~al.}(1999)\citenamefont
  {Gondret}, \citenamefont {Hallouin}, \citenamefont {Lance},\ and\
  \citenamefont {Petit}}]{gondret1999experiments}%
  \BibitemOpen
  \bibfield  {author} {\bibinfo {author} {\bibfnamefont {P.}~\bibnamefont
  {Gondret}}, \bibinfo {author} {\bibfnamefont {E.}~\bibnamefont {Hallouin}},
  \bibinfo {author} {\bibfnamefont {M.}~\bibnamefont {Lance}}, \ and\ \bibinfo
  {author} {\bibfnamefont {L.}~\bibnamefont {Petit}},\ }\bibfield  {title}
  {\enquote {\bibinfo {title} {Experiments on the motion of a solid sphere
  toward a wall: From viscous dissipation to elastohydrodynamic bouncing},}\
  }\href@noop {} {\bibfield  {journal} {\bibinfo  {journal} {Phys. Fluids}\
  }\textbf {\bibinfo {volume} {11}},\ \bibinfo {pages} {2803} (\bibinfo {year}
  {1999})}\BibitemShut {NoStop}%
\bibitem [{\citenamefont {Gondret}\ \emph {et~al.}(2002)\citenamefont
  {Gondret}, \citenamefont {Lance},\ and\ \citenamefont
  {Petit}}]{gondret2002bouncing}%
  \BibitemOpen
  \bibfield  {author} {\bibinfo {author} {\bibfnamefont {P.}~\bibnamefont
  {Gondret}}, \bibinfo {author} {\bibfnamefont {M.}~\bibnamefont {Lance}}, \
  and\ \bibinfo {author} {\bibfnamefont {L.}~\bibnamefont {Petit}},\ }\bibfield
   {title} {\enquote {\bibinfo {title} {Bouncing motion of spherical particles
  in fluids},}\ }\href@noop {} {\bibfield  {journal} {\bibinfo  {journal}
  {Phys. Fluids}\ }\textbf {\bibinfo {volume} {14}},\ \bibinfo {pages} {643}
  (\bibinfo {year} {2002})}\BibitemShut {NoStop}%
\bibitem [{\citenamefont {Joseph}\ \emph {et~al.}(2001)\citenamefont {Joseph},
  \citenamefont {Zenit}, \citenamefont {Hunt},\ and\ \citenamefont
  {Rosenwinkel}}]{joseph2001particle}%
  \BibitemOpen
  \bibfield  {author} {\bibinfo {author} {\bibfnamefont {G.~G.}\ \bibnamefont
  {Joseph}}, \bibinfo {author} {\bibfnamefont {R.}~\bibnamefont {Zenit}},
  \bibinfo {author} {\bibfnamefont {M.~L.}\ \bibnamefont {Hunt}}, \ and\
  \bibinfo {author} {\bibfnamefont {A.~M.}\ \bibnamefont {Rosenwinkel}},\
  }\bibfield  {title} {\enquote {\bibinfo {title} {Particle--wall collisions in
  a viscous fluid},}\ }\href@noop {} {\bibfield  {journal} {\bibinfo  {journal}
  {J. Fluid Mech.}\ }\textbf {\bibinfo {volume} {433}},\ \bibinfo {pages} {329}
  (\bibinfo {year} {2001})}\BibitemShut {NoStop}%
\bibitem [{\citenamefont {Zenit}\ and\ \citenamefont
  {Hunt}(1999)}]{zenit1999mechanics}%
  \BibitemOpen
  \bibfield  {author} {\bibinfo {author} {\bibfnamefont {R.}~\bibnamefont
  {Zenit}}\ and\ \bibinfo {author} {\bibfnamefont {M.~L.}\ \bibnamefont
  {Hunt}},\ }\bibfield  {title} {\enquote {\bibinfo {title} {Mechanics of
  immersed particle collisions},}\ }\href@noop {} {\bibfield  {journal}
  {\bibinfo  {journal} {J. Fluids Eng.}\ }\textbf {\bibinfo {volume} {121}},\
  \bibinfo {pages} {179} (\bibinfo {year} {1999})}\BibitemShut {NoStop}%
\bibitem [{\citenamefont {Barnocky}\ and\ \citenamefont
  {Davis}(1988)}]{barnocky1988elastohydrodynamic}%
  \BibitemOpen
  \bibfield  {author} {\bibinfo {author} {\bibfnamefont {G.}~\bibnamefont
  {Barnocky}}\ and\ \bibinfo {author} {\bibfnamefont {R.~H.}\ \bibnamefont
  {Davis}},\ }\bibfield  {title} {\enquote {\bibinfo {title}
  {Elastohydrodynamic collision and rebound of spheres: experimental
  verification},}\ }\href@noop {} {\bibfield  {journal} {\bibinfo  {journal}
  {Phys. Fluids}\ }\textbf {\bibinfo {volume} {31}},\ \bibinfo {pages} {1324}
  (\bibinfo {year} {1988})}\BibitemShut {NoStop}%
\bibitem [{\citenamefont {Davis}(1987)}]{davis1987elastohydrodynamic}%
  \BibitemOpen
  \bibfield  {author} {\bibinfo {author} {\bibfnamefont {R.~H.}\ \bibnamefont
  {Davis}},\ }\bibfield  {title} {\enquote {\bibinfo {title}
  {Elastohydrodynamic collisions of particles},}\ }\href@noop {} {\bibfield
  {journal} {\bibinfo  {journal} {PhysicoChem. Hydrodyn.}\ }\textbf {\bibinfo
  {volume} {9}},\ \bibinfo {pages} {41} (\bibinfo {year} {1987})}\BibitemShut
  {NoStop}%
\bibitem [{\citenamefont {King}\ \emph {et~al.}(2011)\citenamefont {King},
  \citenamefont {White}, \citenamefont {Maxwell},\ and\ \citenamefont
  {Menon}}]{king2011inelastic}%
  \BibitemOpen
  \bibfield  {author} {\bibinfo {author} {\bibfnamefont {H.}~\bibnamefont
  {King}}, \bibinfo {author} {\bibfnamefont {R.}~\bibnamefont {White}},
  \bibinfo {author} {\bibfnamefont {I.}~\bibnamefont {Maxwell}}, \ and\
  \bibinfo {author} {\bibfnamefont {N.}~\bibnamefont {Menon}},\ }\bibfield
  {title} {\enquote {\bibinfo {title} {Inelastic impact of a sphere on a
  massive plane: Nonmonotonic velocity-dependence of the restitution
  coefficient},}\ }\href@noop {} {\bibfield  {journal} {\bibinfo  {journal}
  {EPL}\ }\textbf {\bibinfo {volume} {93}},\ \bibinfo {pages} {14002} (\bibinfo
  {year} {2011})}\BibitemShut {NoStop}%
\bibitem [{\citenamefont {Flagan}\ and\ \citenamefont
  {Seinfeld}(1988)}]{flagan1988fundamentals}%
  \BibitemOpen
  \bibfield  {author} {\bibinfo {author} {\bibfnamefont {R.~C.}\ \bibnamefont
  {Flagan}}\ and\ \bibinfo {author} {\bibfnamefont {J.~H.}\ \bibnamefont
  {Seinfeld}},\ }\href@noop {} {\emph {\bibinfo {title} {Fundamentals of Air
  Pollution Engineering}}}\ (\bibinfo  {publisher} {Prentice-Hall, Inc.},\
  \bibinfo {year} {1988})\BibitemShut {NoStop}%
\bibitem [{\citenamefont {Ashmore}\ \emph {et~al.}(2005)\citenamefont
  {Ashmore}, \citenamefont {del Pino},\ and\ \citenamefont
  {Mullin}}]{ashmore2005cavitation}%
  \BibitemOpen
  \bibfield  {author} {\bibinfo {author} {\bibfnamefont {J.}~\bibnamefont
  {Ashmore}}, \bibinfo {author} {\bibfnamefont {C.}~\bibnamefont {del Pino}}, \
  and\ \bibinfo {author} {\bibfnamefont {T.}~\bibnamefont {Mullin}},\
  }\bibfield  {title} {\enquote {\bibinfo {title} {Cavitation in a lubrication
  flow between a moving sphere and a boundary},}\ }\href@noop {} {\bibfield
  {journal} {\bibinfo  {journal} {Phys. Rev. Lett.}\ }\textbf {\bibinfo
  {volume} {94}},\ \bibinfo {pages} {124501} (\bibinfo {year}
  {2005})}\BibitemShut {NoStop}%
\bibitem [{\citenamefont {Yang}\ \emph {et~al.}(2006)\citenamefont {Yang},
  \citenamefont {Seddon}, \citenamefont {Mullin}, \citenamefont {Del~Pino},\
  and\ \citenamefont {Ashmore}}]{yang2006motion}%
  \BibitemOpen
  \bibfield  {author} {\bibinfo {author} {\bibfnamefont {L.}~\bibnamefont
  {Yang}}, \bibinfo {author} {\bibfnamefont {J.~R.~T.}\ \bibnamefont {Seddon}},
  \bibinfo {author} {\bibfnamefont {T.}~\bibnamefont {Mullin}}, \bibinfo
  {author} {\bibfnamefont {C.}~\bibnamefont {Del~Pino}}, \ and\ \bibinfo
  {author} {\bibfnamefont {J.}~\bibnamefont {Ashmore}},\ }\bibfield  {title}
  {\enquote {\bibinfo {title} {The motion of a rough particle in a stokes flow
  adjacent to a boundary},}\ }\href@noop {} {\bibfield  {journal} {\bibinfo
  {journal} {J. Fluid Mech.}\ }\textbf {\bibinfo {volume} {557}},\ \bibinfo
  {pages} {337} (\bibinfo {year} {2006})}\BibitemShut {NoStop}%
\bibitem [{\citenamefont {Landau}\ and\ \citenamefont
  {Lifshitz}(1986)}]{landau1986theory}%
  \BibitemOpen
  \bibfield  {author} {\bibinfo {author} {\bibfnamefont {L.~D.}\ \bibnamefont
  {Landau}}\ and\ \bibinfo {author} {\bibfnamefont {E.~M.}\ \bibnamefont
  {Lifshitz}},\ }\href@noop {} {\emph {\bibinfo {title} {Theory of
  Elasticity}}}\ (\bibinfo  {publisher} {Elsevier New York},\ \bibinfo {year}
  {1986})\BibitemShut {NoStop}%
\bibitem [{\citenamefont {Leal}(2007)}]{leal2007advanced}%
  \BibitemOpen
  \bibfield  {author} {\bibinfo {author} {\bibfnamefont {L.~G.}\ \bibnamefont
  {Leal}},\ }\href@noop {} {\emph {\bibinfo {title} {Advanced Transport
  Phenomena: Fluid Mechanics and Convective Transport Processes}}}\ (\bibinfo
  {publisher} {Cambridge University Press},\ \bibinfo {year}
  {2007})\BibitemShut {NoStop}%
\end{thebibliography}%

\end{document}